# Speeding up of microstructure reconstruction:
# I. Application to labyrinth patterns

**R Piasecki and W Olchawa**

Institute of Physics, University of Opole, Oleska 48, 45-052 Opole, Poland

E-mail: piaser@uni.opole.pl

**Abstract**

Recently, entropic descriptors based the Monte Carlo hybrid reconstruction of the microstructure of a binary/greyscale pattern has been proposed (Piasecki 2011 *Proc. R. Soc.* A **467** 806). We try to speed up this method applied in this instance to the reconstruction of a binary labyrinth target. Instead of a random configuration, we propose to start with a suitable synthetic pattern created by cellular automaton. The occurrence of the characteristic attributes of the target is the key factor for reducing the computational cost that can be measured by the total number of MC steps required. For the same set of basic parameters, we investigated the following simulation scenarios: the biased/random alternately mixed #2m approach, the strictly biased #2b and the random/partially biased #2rp one. The series of 25 runs were performed for each scenario. To maintain comparable accuracy of the reconstructions, during the final stages the only selection procedure we used was the biased one. This allowed us to make the consistent comparison of the first three scenarios. The purely random #2r approach of low efficiency was included only for completeness of the approaches. Finally, for the conditions established, the best single reconstruction and the best average tolerance value among all the scenarios were given by the mixed #2m method, which was also the fastest one. The slightly slower the alternative #2b and #2rp variants provided comparable but less satisfactory results.

(Some figures may appear in colour only in the online journal)

## 1. Introduction

For efficient Monte Carlo (MC) modelling, a reduction in the reconstruction time is highly desirable. Can we reduce it considerably in a simple way, at least for a certain class of patterns? Let us focus on the microstructure reconstruction of a binary labyrinth pattern making use of the simulated annealing (SA) technique within the MC method. There are many factors that influence the computational cost—measured roughly by the total number of MC steps (MCSs)—in particular, when a personal computer is in use. In this context, a few essential factors are addressed in this paper. Frequently, a random initial configuration of objects is applied instead of a suitable synthetic pattern that includes some of the most characteristic attributes of a target. To minimize an objective function, the assumed level of the tolerance value should not be too restrictive. We mean such a situation when it nearly doubles the number of MCSs while the final pattern is practically of the same quality. In addition, the optimization of a used program and the choice of a kind of simulation scenario belong to the essential factors. Each of these points is taken into account in this paper, with a particular care regarding the first and last ones.

Among the numerous models of pattern formation, the cellular automata (CA) approach is particularly simple in usage. To this group belongs spatially discrete model of vertebrate skin pattern growth proposed by Young [1]. The mechanism of pattern formation is based on short-range activation and long-range inhibition. The model allows the reproduction of the basic features of vertebrate skin patterns: spots, stripes and mixed forms. Its extended version (given in the next section) can be applied to produce a synthetic initial pattern with the target's characteristic attributes. In this way, the microstructure reconstruction becomes more efficient. It should also be noted that our aim is not the exact duplication of the parent microstructure. The inverse reconstruction process is meant rather to create *statistically* similar microstructures instead [2, 3]. Such microstructures can be used for testing of various models or to optimise the design of heterogeneous materials in material science [4, 5].

Unlike the traditionally used correlation functions [6–9], objective functions employing the so-called entropic descriptors (EDs) have been recently proposed [10]. A different fresh approach that describes microstructure variance as a stochastic process was developed in [11]. In this paper, we focus on the ED method that makes multi-scale information known partly dissimilar to that given by correlation functions at almost all length scales. In the simplest case, the EDs can be used for quantitative characterization of the spatial inhomogeneity of a binary microstructure (for greyscale images other applications are also available [10]). Here we show, how the ED based MC hybrid reconstruction of labyrinth binary microstructure can be made much faster in a simple way.

## 2. The extended Young's model

According to specific rules of Young's model [1], an initially random arrangement of $n_{init}$ differentiated cells (DC as black pixels) in a matrix of undifferentiated cells (UC as white pixels) can evolve into a 'skin' pattern. A square grid $L \times L$ with periodic boundary conditions in both directions is used for simulation of the evolution. Each DC produces two diffusible morphogens at a constant rate: i) an inhibitor having the longer range and stimulating the dedifferentiation of DC, ii) an activator stimulating the differentiation of the nearby UC. On the other hand, the UCs are passive and produce no active substances. The sum of morphogens influences each $(x, y)$-cell from all neighbouring DCs and this determines the fate of the cell. Therefore, in this process the mechanism of pattern formation includes local activation and long-range inhibition:

$$w(r_i) = \begin{cases} w_1 > 0 & \text{for} \quad r_i \leq R_1 \quad (I - \text{region}), \\ w_2 < 0 & \text{for } R_1 < r_i \leq R_2 \quad (II - \text{region}). \end{cases} \tag{2.1}$$

Here, $r_i$ is the radial distance of the $i$th DC from the $(x, y)$-cell. For the model parameters $w_1$, $w_2$, $R_1$ and $R_2$, the rule of time-evolution of every cell, see (2.3), depends on the summary activation-inhibition field at time $t$

$$W(x, y; t) = \sum_{r_i \in I} w_1 + \sum_{r_i \in II} w_2 \tag{2.2}$$

Within a simple extension of the above model, the process of evolution affected by chemical or physical properties of an 'environment' can be easily taken into account. Making use of the auxiliary parameter $\varepsilon$ that is involved in on−off switching of cell differentiations, favourable ($\varepsilon < 0$), neutral ($\varepsilon = 0$) or hostile ($\varepsilon > 0$) environmental conditions can be considered. Thus, for each $(x, y; t)$-cell the following situations can appear at the time $t + 1$:

(a)  If  $W(x, y; t) < \varepsilon$  then DC(UC) becomes (remains)  a UC cell at time  $t + 1$

(b)  If  $W(x, y; t) = \varepsilon$  then the cell does not changes state at time  $t + 1$   (2.3)

(c)  If  $W(x, y; t) > \varepsilon$  then UC(DC) becomes (remains)  a DC cell at time  $t + 1$

If $\varepsilon > 0$ then we observe the lowering of $n_{\text{final}}$ in comparison with the original model that is recovered for $\varepsilon = 0$. The opposite situation appears for $\varepsilon < 0$.

Once the results of changing states for each grid cell are saved as a separate successive pattern, we consider this moment as the first iteration step $j = 1$ (temporal evolution step). Then, the resulting black−white pattern with a current population size $n(j)$ becomes the new starting configuration. Let us denote the number of 'positive' UC → DC and 'negative' DC → UC changes in the $j$th iteration step by $\Delta n^+(j)$ and $\Delta n^-(j)$, respectively. The iteration is repeated until the resulting synthethic pattern no longer changes, i.e. $\Delta n^+(j) = \Delta n^-(j) = 0$  [1].

### 3. Hybrid approach to microstructure reconstruction

The question of modelling heterogeneous materials—'to what extent can the structure of a disordered heterogeneous material be reconstructed using limited but essentially exact structural information about the original system?'—remains a present-day problem [12]. Here we apply one of the particularly useful approaches of modelling in material science, the SA technique within Monte Carlo method; see for instance the monograph [13]. In this paper, a $100 \times 100$ sub-domain of a larger binarized image of magnetic stripes in perpendicular Co/Pt-based multilayers, see figure 2 in [14], was adapted as the target pattern. This magnetic force microscopy image corresponds to a quite complex labyrinth microstructure with the black phase volume fraction $\varphi = 0.591$. In order to perform its statistical reconstruction we consider the multi-scale function $E$ fully described in [10]. This objective function is the weighted sum of squared differences between EDs computed for trial configurations and targets ones marked with superscript '0'

$$E = \frac{1}{4} \sum_{k=1}^{L} [(S_\Delta - S_\Delta^0)^2 + (C_{\lambda,S} - C_{\lambda,S}^0)^2 + (G_\Delta - G_\Delta^0)^2 + (C_{\lambda,G} - C_{\lambda,G}^0)^2]. \tag{3.1}$$

The basic definition of the EDs is given below in (3.2) and (3.3). In particular, the ED $S_\Delta$ ($C_{\lambda,S}$) quantifies the *spatial* inhomogeneity (statistical complexity) [15, 16], while $G_\Delta$ ($C_{\lambda,G}$) describes the *compositional* inhomogeneity (statistical complexity), respectively [16, 17]. We recall that a binary pattern can be encoded in two ways: (a) standard one (0 =black, 1 =white) and (b) greyscale fashion (0 =black, 255 =white). Therefore, to increase the statistical sensitivity of the reconstruction, in (3.1) we employ ED pairs: $\{S_\Delta, C_{\lambda,S}\}$ and $\{G_\Delta, C_{\lambda,G}\}$. Each of the EDs makes use of the microcanonical entropy $Entr(X) = k_B \ln \Omega(X)$, where Boltzmann's constant is equal to unity, $\Omega$ denotes the number of microstates realizing a macrostate properly defined and $X = \{S$ for the standard binary case; $G$ for the greyscale case$\}$.

In our approach, black pixels are treated as finite-sized $1 \times 1$-objects. A configurational (binary) macrostate is given by a set of numbers $\{n_i(k)\}$, $i = 1, 2, \ldots, \lambda(k)$, of black pixels inside the cell of size $k \times k$ sliding by a discrete unit distance. The side length of the cell defines the length scale $k$. For a one-dimensional pattern, $1 \times L$, the number of allowed positions of the sliding cell $1 \times k$ is equal to $\lambda(k) = [L - k + 1]$ while in a two-dimensional case, $L \times L$, we have the sliding cell $k \times k$ and $\lambda(k) = [L - k + 1]^2$ possible locations. Note that instead of letters $\kappa$ and $\chi$ previously used in [10, 15], now we make use of $\lambda$ exclusively.

In turn, a set of cell sums $\{g_i(k)\}$, $i = 1, 2, \ldots, \lambda(k)$, of grey level values determines a compositional (greyscale) macrostate. The more detailed description and formulae used for the numbers of realizations $\Omega$ of appropriate macrostates can be found in the appendix of [10].

For inhomogeneity, the general form of EDs $S_\Delta$ and $G_\Delta$ is as follows:

$$X_\Delta(k) = \frac{1}{\lambda} \left[ Entr_{max}(X) - Entr(X) \right] \ . \tag{3.2}$$

The highest possible value of entropy, $Entr_{max}(X; k) = \ln \Omega_{max}(X; k)$, is related to the most spatially (or compositionally) uniform object configuration at a given length scale. Such an arrangement of black pixels (or distribution of grey levels) corresponds to the reference macrostate $RM_{max}(X)$. To evaluate, for each length scale $k$, the deviation of the current arrangement that represents the actual macrostate AM from the appropriate $RM_{max}$ it is natural to consider the difference $Entr_{max}(k) - Entr(k)$ averaged over the number of cells $\lambda$. The averaging allows for a comparison of ED values at different length scales $k$.

On the other hand, for statistical complexity $C_{\lambda, S}$ and $C_{\lambda, G}$ have a more extended form

$$C_{\lambda, X}(k) = \frac{1}{\lambda} \frac{[Entr_{max}(X) - Entr(X)][Entr(X) - Entr_{min}(X)]}{[Entr_{max}(X) - Entr_{min}(X)]} \ . \tag{3.3}$$

Here, the lowest possible value of entropy, $Entr_{min}(X; k) = \ln \Omega_{min}(X; k)$, is related to the most spatially (or compositionally) non-uniform object configuration at a given length scale. For limiting macrostates characterised by $Entr_{max}(k)$ and $Entr_{min}(k)$, the statistical complexity should be minimal and it is, see (3.3). The most 'complex' arrangement at a given length scale emerges when the average *departures* of the actual entropy from the highest one and from the lowest reference entropy are similar to each other, a kind of compromise between two opposite limiting configurations: the most homogeneous and the most inhomogeneous.

According to (3.2) for any length scale $1 \leq k \leq L$, the binary $S_\Delta$ component (the grey level counterpart $G_\Delta$) of the ED pairs takes into account the statistical *dissimilarity* of the actual macrostate AM(S) [AM(G)] and the reference theoretical one $RM_{max}(S)$ [$RM_{max}(G)$] that maximizes the appropriate entropy. Consecutively, in (3.3) the binary $C_{\lambda, S}$ component of the ED pairs takes into consideration the statistical *dissimilarity* of macrostates in the pairs:

AM($S$) and RM$_{max}$($S$), AM($S$) and RM$_{min}$($S$), RM$_{max}$($S$) and RM$_{min}$($S$), where the reference theoretical one RM$_{min}$($S$) [RM$_{min}$($G$)] minimizes the appropriate entropy. A similar interpretation applies to the grey level counterpart $C_{\lambda, G}$. This type of ED is able to distinguish structurally distinct configurational [compositional] macrostates with identical or nearly the same degree of spatial [compositional] disorder [16]. In particular, we are interested in those structural features that depend on the length scale $k$.

To minimize the objective function $E$, see (3.1), we repeat the following steps. For a current pattern, two randomly selected pixels of different phases are interchanged giving the new trial configuration. The new 'state' is then accepted with probability $p(\Delta E)$ given by the Metropolis–MC acceptance rule [13]

$$p(\Delta E) = \begin{cases} 1 & \Delta E \leq 0, \\ \exp(-\Delta E/T), & \Delta E > 0, \end{cases} \tag{3.4}$$

where $\Delta E = E_{new} - E_{old}$ is the change in the 'energy' between the two successive states. Upon acceptance, the trial pattern becomes a current one, and the evolving procedure is repeated. The variation of a fictitious temperature $T$ as a function of time is called the cooling schedule associated with the annealing process. We use the popular cooling schedule $T(l)/T(0) = \gamma^l$ with a fixed parameter $\gamma = 0.8$, where $l$ stands for the number of temperature loops. For each $l$, the value of the number $n_a(l)$ of MCS accepted during the $l$th temperature loop allows for tracing a fraction $n_a(l)/N(l) \equiv \alpha(l)$, where $N(l)$ means the entire length of the $l$th loop. In turn, the chosen value of initial temperature, $T(0) = 4 \times 10^{-3}$, ensures the initial effectiveness $0.3 < \alpha(1) < 0.65$ for the synthetic starting pattern '2' that possesses characteristic attributes of the target's pattern T; see the corresponding insets in figure 1. When a fraction of less than 1% ($\alpha < 0.01$) of the proposed MCS is accepted, we put in motion the specified biased procedure of exchange of positions of two pixels of different phases. Such a pair is randomly selected among the black pixels belonging to a black cluster's border and white pixels neighbouring the border.

It should be stressed that the biased procedure consists of the conditional acceptation of the pixels exchange. It is carried out, when the number of nearest neighbour (nn) black pixels for a white centre is greater than or equal to number two and simultaneously, the number of nn black pixels for a black centre is less than or equal to number two. As a result, the number of very small white clusters (inside the black 'sea') and black clusters (inside the white 'sea') reduces considerably. Two pixels aligned along the diagonal form two clusters each of size one while their side contact forms one cluster of size two. Additionally, some of the biased exchanges on a cluster's surface modify their shapes. This is in contrast to the strictly unbiased multi-scale reconstruction used previously [10], where the target microstructure was naturally composed of many inner clusters of sizes in a wide range. Here, we deal with a specific labyrinth pattern. There, in addition to one compact large cluster of size 5789 in pixels, only a few much smaller clusters appear. Their occurrence is caused mainly by the procedure of cutting out a $100 \times 100$ target sub-domain from the larger pattern adapted from [14]. This is an additional reason for the application of a biased procedure.

## 4. Four simulation scenarios

The Monte Carlo reconstructions typically use a random configuration as the initial one. This contributes to an increase in the computational cost. At least for a certain class of patterns, i.e. labyrinth patterns, this point can be easily overcome. To reduce the computational cost of the statistical reconstruction process, we propose to replace a random initial configuration '1' (not shown) by the synthetic pattern '2'; see the inset of figure 1. The latter pattern can be easily created with the CA program already discussed in section 2. For a given seed, it is enough to take into account just two main indicators, the position of the first peak of $S_\Delta$ placed at length scale $k = 5$ and the number $n_{\text{final}} = 5789$ of black pixels corresponding to its volume fraction $\varphi = 0.591$. Having some experience with the CA program, several trials led to the following model parameters: $R_1 = 1.8$, $R_2 = 6.2$, $w_1 = 1$, $w_2 = -0.068$, $\varepsilon = -0.127$. Also the initial number of black pixels, $n_{\text{init}} = 493$, can be finally fitted. In figure 1, the $S_\Delta$ lines, dashed and solid (red online), show how close the microstructure of pattern '2' is to the target one T.

Usually, the evolving procedure ends when the energy $E$ becomes smaller than the assumed tolerance $\delta$-value. We think that a better insight into the efficiency of different simulation strategies can be gained by the comparison of results for a series of simulations under clearly established conditions. Here, we ask how far from the chosen value $\delta = 3 \times 10^{-3}$ are the average results for each of the proposed scenarios with the limited number of loops, $l_{\text{max}} = 32$, but with increasing length of successive $l$th loop. The length $N(l)$ of each of the loops is an integer part of the suggested function $f(l)$ given by

$$f(l) = a + (l-1)b + (c^{\frac{l-1}{l^*}} - 1) , \qquad (4.1)$$

where $l = 1, 2, ..., l_{max}$ is the consecutive number of a temperature loop. For higher $l$ numbers than the chosen one, $l^* = 20$ in (4.1), the length increase becomes more and more non-linear; see the inset of figure 2. The other coefficients are $a = 100$, $b = 25$, $c = 750$ and correspond to the length of initial loop, the linear part of the increase in loop length and the base of power function, respectively. In this way, for different scenarios comparable loop parameters are ensured.

It should be also mentioned that the present program is significantly optimized. Namely, for calculation of the factorial function, particularly at large arguments, we apply the code taken from [18], which is based on the excellent approximation for the gamma function derived by Lanczos [19]. Compared with the version used previously this code works very fast and produces highly accurate results.

Now, using the synthetic initial pattern '2', we investigate three main simulation scenarios: the mixed one #2m with alternately applied biased/random selection in the succeeding temperature loops, the strictly biased #2b and the random/partially biased (more precisely, biased only at the final stage) #2rp. Additionally, for comparison purposes only, the purely random #2r approach with no use of the biased procedure is also considered. We would like to point out that a random initial configuration '1' is not taken into account, since such an approach is highly inefficient. It needed about 20 000 acceptable MCS during a single long run of the previous version of our program.

Among the tested variants, the mixed one #2m provides the best results. In this approach, the length of those temperature loops, which were involved in a random selection procedure, was shortened to about 1/5 of the $l$th length described by (4.1). The best #2m single reconstruction among the 25 runs is also the best one in comparison with the other scenarios performed for the same conditions. It shows the tolerance value $\delta = 2.5 \times 10^{-3}$ obtained after $N_a = 1266$ accepted and $N_{tot} = 113\ 320$ total MCS; see (red) circles and the upper right inset of figure 1. In this figure, the $S_\Delta$ thick solid line corresponds to the labyrinth target T-pattern, short dashed (red) line relates to the synthetic '2' initial configuration created suitably by cellular automaton described in section 2. In turn, the long-dashed (blue) line refers to $S_\Delta$ calculated for a random initial configuration '1' (not used further in fact). The appearance of equal quality fitting was also found for the other three EDs (not shown here for clarity of the figure). The adequate statistical features of the best #2m reconstruction are also confirmed by the cluster size distribution shown in table 1. In addition, one can see therein that the smallest average tolerance value $<\delta> = 3.46 \times 10^{-3} \pm 0.000\ 36$ belongs to #2m scenario.

**Table 1.** Numbers $n(s)$ of cluster sizes $s$ (in pixels) for target pattern T and its the best representative reconstructions #2m, #2b, #2rp and #2r, compared for the same conditions. Below, the corresponding patterns are depicted. The quality of reconstructions can be measured by the tolerance $\delta$-value. For scenarios investigated, the average $<\delta>$ of 25 runs of our main program is given. In addition, the number $N_a$ of acceptances and the number $N_{tot}$ of MCS during 32 loops of a single run is shown. The last right pattern and shadowed data relate to the extra-reconstruction of type #2m, when only $\gamma \to 0.85$ was changed.

| T | | #2m | | #2b | | #2rp | | #2r | | continuation | |
|---|---|---|---|---|---|---|---|---|---|---|---|
| $s$ | $n$ | $s$ | $n$ | $s$ | $n$ | $s$ | $n$ | $s$ | $n$ | | |
| 5 | 1 | 1 | 6 | 1 | 1 | 1 | 15 | 1 | 57 | 69 | 1 |
| 8 | 1 | 2 | 1 | 3 | 1 | 29 | 1 | 2 | 5 | 77 | 1 |
| 10 | 1 | 37 | 1 | 55 | 1 | 35 | 1 | 3 | 2 | 84 | 1 |
| 22 | 1 | 62 | 1 | 5851 | 1 | 50 | 1 | 5 | 1 | 108 | 1 |
| 36 | 1 | 92 | 1 | | | 56 | 1 | 9 | 1 | 129 | 1 |
| 40 | 1 | 5711 | 1 | | | 125 | 1 | 18 | 1 | 348 | 1 |
| 5789 | 1 | | | | | 5600 | 1 | 34 | 1 | 539 | 1 |
| | | | | | | | | 44 | 1 | 602 | 1 |
| | | | | | | | | 48 | 1 | 805 | 1 |
| | | | | | | | | 51 | 3 | 2708 | 1 |
| | | | | | | | | 57 | 1 | | |

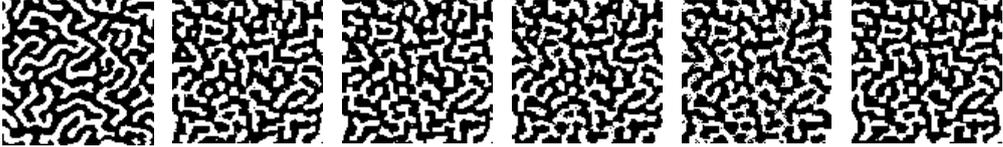

#2m for $\gamma \to 0.85$
with $s_{max} = 5815$

| | | | | | |
|---|---|---|---|---|---|
| $\delta$ = 0.00250 | 0.00342 | 0.00306 | 0.03002 | 0.00201 |
| $<\delta>$ = 0.00346 | 0.00409 | 0.00435 | 0.03419 | |
| ± 0.00036 | ± 0.00041 | ± 0.00054 | ± 0.00435 | |
| $N_a \cong$ 1266 | 1562 | 1375 | 1061 | 2331 |
| $<N_a> \cong$ 1324 | 1612 | 1341 | 1047 | |
| $N_{tot}$ = 113320 | 117054 | 117054 | 117054 | 106340 |

In the same table, the data corresponding to the best representatives of other scenarios are also summarized. For instance, the size of the biggest cluster $s$(#2m) = 5711 is quite close to the target one $s$(T) = 5789 similarly to the strictly biased approach, $s$(#2b) = 5851. However, the latter one does not guarantee a suitable distribution of the cluster's sizes. Similar to the #2b case behaviour for a distribution of the cluster's sizes appears for the #2rp scenario; see table 1. In the latter case the random selection ends in the $l$th loop if $\alpha(l) < 1$ and the biased selection begins in the $(l+1)$th loop until $l = l_{max} = 32$. Despite such disadvantages, only slightly worse average tolerance values, $<\delta$(#2b)$> = 4.09 \times 10^{-3} \pm 0.00041$, and $<\delta$(#2rp)$> = 4.35 \times 10^{-3} \pm 0.00054$, are shown by these scenarios. The best reconstruction of 25 runs for the former approach, needed $N_a$(#2b) = 1562 accepted and $N_{tot}$ = 117 054 of MCS while the latter one, only $N_a$(#2rp) = 1375. Therefore, both #2b and #2rp scenarios show moderate advantages as well as weaknesses. In this comparison, the fastest #2m scenario, with its average (over 25 runs) total number $<N_{tot}> \cong 111\,501$ of MCS is our preferred approach.

**Table 2.** Numbers $n_a(l)$ and fractions $\alpha(l)$ of accepted trials in each loop for the best representative reconstructions #2m, #2b, #2rp and #2r obtained for the same conditions. In the last two sub-columns, the similar data are shown for the extra reconstruction of type #2m, when only $\gamma \to 0.85$ was changed. The numbers of acceptances for non-biased trials are marked in bold.

| | #2m | | #2b | | #2rp | | #2r | | #2m for $\gamma \to 0.85$ | |
|---|---|---|---|---|---|---|---|---|---|---|
| $l$ | $n_a(l)$ | $\alpha(l)$ | $n_a(l)$ | $\alpha(l)$ | $n_a(l)$ | $\alpha(l)$ | $n_a(l)$ | $\alpha(l)$ | $n_a(l)$ | $\alpha(l)$ |
| 1 | 57 | 57 | 63 | 63 | **32** | 32 | **35** | 35 | 59 | 59 |
| 2 | **10** | 47.62 | 64 | 51.2 | **23** | 18.4 | **26** | 20.8 | **8** | 38.1 |
| 3 | 72 | 47.68 | 74 | 49.01 | **18** | 11.92 | **26** | 17.22 | 78 | 51.66 |
| 4 | **7** | 24.14 | 76 | 42.94 | **23** | 12.99 | **29** | 16.38 | **14** | 48.28 |
| 5 | 83 | 40.89 | 70 | 34.48 | **18** | 8.87 | **26** | 12.81 | 108 | 53.20 |
| 6 | **4** | 10.53 | 79 | 34.5 | **19** | 8.3 | **17** | 7.42 | **7** | 18.42 |
| 7 | 85 | 33.20 | 91 | 35.55 | **22** | 8.59 | **12** | 4.69 | 116 | 45.31 |
| 8 | **8** | 17.02 | 72 | 25.35 | **25** | 8.80 | **14** | 4.93 | **9** | 19.15 |
| 9 | 83 | 26.52 | 75 | 23.96 | **19** | 6.07 | **17** | 5.43 | 111 | 35.46 |
| 10 | **8** | 14.04 | 74 | 21.51 | **16** | 4.65 | **20** | 5.81 | **6** | 10.53 |
| 11 | 80 | 21.28 | 73 | 19.41 | **17** | 4.52 | **20** | 5.32 | 99 | 26.33 |
| 12 | **4** | 5.80 | 71 | 17.23 | **9** | 2.18 | **18** | 4.37 | **13** | 18.84 |
| 13 | 79 | 17.48 | 62 | 13.72 | **21** | 4.65 | **10** | 2.21 | 115 | 25.44 |
| 14 | **6** | 6 | 68 | 13.65 | **21** | 4.22 | **25** | 5.02 | **10** | 10 |
| 15 | 64 | 11.59 | 55 | 9.96 | **21** | 3.80 | **13** | 2.36 | 109 | 19.75 |
| 16 | **4** | 3.25 | 42 | 6.81 | **25** | 4.05 | **21** | 3.40 | **9** | 7.317 |
| 17 | 78 | 11.16 | 46 | 6.58 | **26** | 3.72 | **23** | 3.29 | 127 | 18.17 |
| 18 | **5** | 3.13 | 47 | 5.86 | **18** | 2.24 | **19** | 2.37 | **6** | 3.75 |
| 19 | 84 | 8.97 | 44 | 4.70 | **29** | 3.1 | **29** | 3.1 | 130 | 13.89 |
| 20 | **1** | 0.45 | 35 | 3.14 | **29** | 2.61 | **38** | 3.41 | **6** | 2.69 |
| 21 | 68 | 5.04 | 26 | 1.92 | **38** | 2.82 | **28** | 2.08 | 108 | 8.01 |
| 22 | 62 | 3.72 | 38 | 2.28 | **38** | 2.28 | **22** | 1.32 | **5** | 1.5 |
| 23 | 59 | 2.81 | 22 | 1.05 | **39** | 1.85 | **44** | 2.09 | 115 | 5.47 |
| 24 | 25 | 0.93 | 33 | 1.22 | **39** | 1.45 | **36** | 1.34 | **8** | 1.48 |
| 25 | 37 | 1.05 | 31 | 0.88 | **68** | 1.93 | **55** | 1.56 | 143 | 4.06 |
| 26 | 28 | 0.60 | 19 | 0.41 | **66** | 1.42 | **52** | 1.12 | **4** | 0.34 |
| 27 | 35 | 0.56 | 20 | 0.32 | **50** | 0.80 | **49** | 0.79 | 168 | 2.70 |
| 28 | 19 | 0.23 | 24 | 0.29 | 361 | 4.31 | **52** | 0.62 | 121 | 1.44 |
| 29 | 33 | 0.29 | 19 | 0.17 | 108 | 0.95 | **61** | 0.54 | 170 | 1.49 |
| 30 | 27 | 0.17 | 17 | 0.11 | 43 | 0.28 | **85** | 0.55 | 124 | 0.8 |
| 31 | 28 | 0.13 | 22 | 0.10 | 38 | 0.18 | **62** | 0.29 | 102 | 0.48 |
| 32 | 23 | 0.08 | 10 | 0.03 | 56 | 0.19 | **77** | 0.26 | 123 | 0.42 |

As concerns the entirely random scenario #2r, it was included here only for completeness of investigation. Note that in this case the average tolerance value, $<\delta(\#2r)> = 3.42 \times 10^{-2} \pm 0.004\,35$, is relatively worse than the values earlier discussed by about one order. This weak result is expected since the purely random selection procedure from a definition is not well suited to optimize the cost MC function for a highly correlated initial synthethic pattern composed of compact clusters. We refer the reader to tables 1 and 2, where various details of different scenarios are summarized.

In order to obtain a better insight into the speeding up of ED-based microstructure reconstruction, in table 2 the numbers $n_a(l)$ and fractions $\alpha(l)$ of accepted trials in each temperature loop for the best representative reconstructions #2m, #2b, #2rp and #2r are summarized. It is interesting that in the sub-column related to the #2rp method, the jump of $n_a(l)$ appears for $l=28$, since the random selection was stopped earlier and the biased exchange procedure is switched on. Such a switching at the final stage is forbidden for purely random #2r scenario, hence, the worse results appear.

One point deserves a comment; in both tables, quite surprising data are presented for a single run of extra-reconstruction of type #2m but with only one parameter changed, i.e. $\gamma \to 0.85$. The less aggressive cooling schedule results in the best tolerance value $\delta = 2.01 \times 10^{-3}$ that was obtained after $N_{a,\,extra} = 2331$ of accepted MCS; see the shadowed data and the corresponding final pattern in table 1. Moreover, instead of temperature $T(l=32) = 3.96 \times 10^{-6}$ now the program ends with nearly 7 times higher temperature, $T_{extra}(l=32) = 2.59 \times 10^{-5}$. This also forces the appearance of higher acceptance numbers $n_a(l)$ during each temperature loop and their more regular distributions; see the last two sub-columns in table 2. Finally, this single run turned needed only $N_{tot,\,extra} = 106\,340$ of MCS. All these indicate that there is still much room for searching optimal scenarios, at least for the reconstruction of binary labyrinth patterns.

Figure 2 illustrates the convergence of the cost function $E$ for different scenarios we have focused on. The arrows indicate the ED-based reconstructions: #2m (red line), #2b (green line), #2rp (blue line) and #2r (grey line) performed under the same conditions. In addition, the convergence for extra-reconstruction of type #2m is shown, when only $\gamma \to 0.85$ is changed (rose line). One can also observe the changes of the number $N_a$ of accepted MCS for each of the investigated scenarios. It is obvious now that the smallest acceptance number cannot be linked directly with the best quality reconstruction, see for instance, the grey line corresponding to the best representative reconstruction of #2r type.

Now, we would like to point out that an interesting alternative [20] for reconstructing random microstructures is the hybrid approach making use of a pair $\{S_2(r);\, C_2(r)\}$, where two-point cluster function $C_2(r)$ gives the probability of finding two points separated by a distance $r$ in the same cluster of the phase of interest. Figure 3 displays the restricted two-point correlation function $S_2$ of the black phase computed by orthogonal-sampling algorithm with the hard-wall conditions for the target T-pattern (black line) and for the best reconstructions #2m (red circles), #2b (green triangles), #2rp (blue squares). Those reconstructions are of comparable fitting quality because of the similar $\delta$ except the next one, #2r (grey line). One can observe that even such a simplified variant of two-point correlation function reveals relatively dissimilar structural information from that given by EDs for all reconstructed patterns. For instance, see the lower amplitudes at important length scale $k = 5$ and stronger fluctuations at larger scales, say for $k > 15$. Therefore, the combination of our method with the standard two-point correlation function $S_2$ may be interesting in this context.

## 4. Summary

We have developed a simple approach that allowed for speeding up of microstructure reconstruction of binary (two phases) labyrinth patterns. The effectiveness of the replacing of a random initial configuration with a synthetic pattern carefully prepared by cellular automaton has been confirmed by a series of tests. The hybrid reconstruction based on novel entropic descriptors seems to be a quite prospective method. It provides a responsible statistical reconstruction of a microstructure with a given tolerance value. The number of accepted and total MCS reduces significantly also as result of the use of an improved version of our program. It employs only 32 temperature loops of given different lengths. For the established conditions, the most cost-effective scenario leading simultaneously to the best quality of the reconstruction is the biased/random alternately mixed #2m approach. From the physical point of view, such an outcome agrees well with our expectations.

## Figure captions

**Figure 1.** Hybrid reconstruction making use of two pairs of EDs, $\{S_{\Delta}; C_{\lambda, S}\}$ and $\{G_{\Delta}; C_{\lambda, G}\}$, for a $100 \times 100$ part of the binarized image of magnetic stripe domains in perpendicular Co/Pt-based multilayers adapted from figure 2 in [10]. The $S_{\Delta}$ thick solid line corresponds to target labyrinth T-pattern while fitting by red circles refers to the mixed-type #2m reconstruction; the short dashed red line relates to the synthetic '2' initial configuration created suitably by a cellular automaton while the long dashed blue line indicates the $S_{\Delta}$ calculated for a random initial configuration. Equal quality fitting appears also for the other three EDs.

**Figure 2.** Convergence of different preferred scenarios for EDs based hybrid reconstructions: #2m (red line), #2b (green line), #2rp (blue line) and #2r (grey line) performed for the same conditions. In addition, the convergence for extra reconstruction of type #2m is shown, when only $\gamma \rightarrow 0.85$ is changed (rose line). In the inset, the increase in loop length as a function of loop number $l$ is depicted.

**Figure 3.** Restricted two-point correlation function $S_2$ of black phase computed by the orthogonal-sampling algorithm with the hard-wall conditions for target T-pattern (black line) and for its reconstructions: #2m (red circles), #2b (green triangles), #2rp (blue squares) and #2r (grey line). The easily seen differences between the target and its reconstructions (via EDs) show that they provide relatively dissimilar structural information compared with $S_2$.

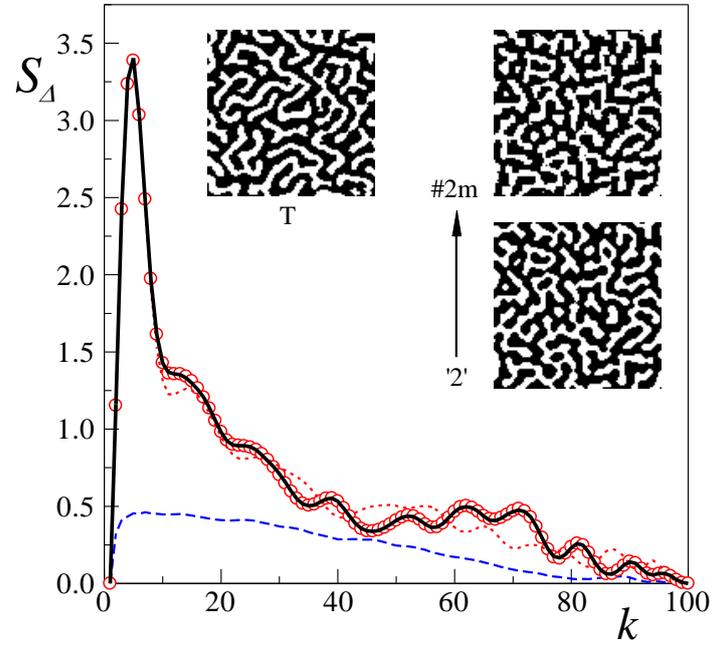

**Figure 1.** Hybrid reconstruction making use of two pairs of EDs, $\{S_{\Delta}; C_{\lambda, S}\}$ and $\{G_{\Delta}; C_{\lambda, G}\}$, for a $100 \times 100$ part of the binarized image of magnetic stripe domains in perpendicular Co/Pt-based multilayers adapted from figure 2 in [10]. The $S_{\Delta}$ thick solid line corresponds to target labyrinth T-pattern while fitting by red circles refers to the mixed-type #2m reconstruction; the short dashed red line relates to the synthetic '2' initial configuration created suitably by a cellular automaton while the long dashed blue line indicates the $S_{\Delta}$ calculated for a random initial configuration. Equal quality fitting appears also for the other three EDs.

Fig. 1

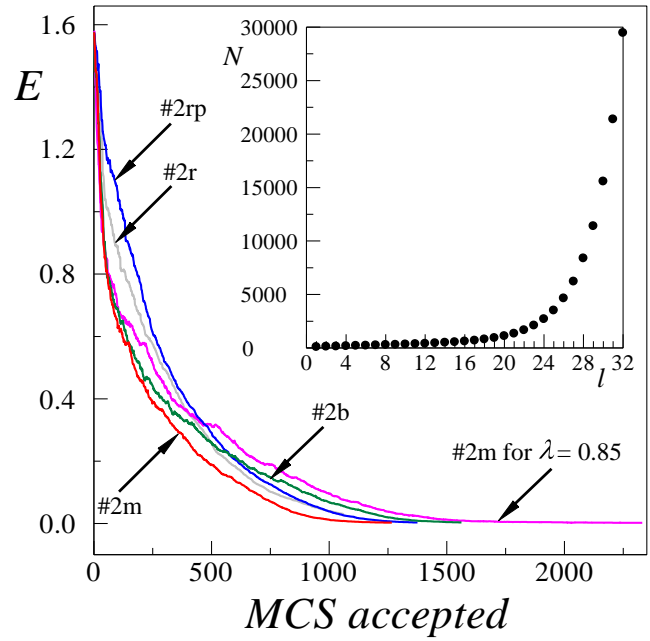

**Figure 2.** Convergence of different preferred scenarios for EDs based hybrid reconstructions: #2m (red line), #2b (green line), #2rp (blue line) and #2r (grey line) performed for the same conditions. In addition, the convergence for extra reconstruction of type #2m is shown, when only $\gamma \rightarrow 0.85$ is changed (rose line). In the inset, the increase in loop length as a function of loop number $l$ is depicted.

Fig. 2

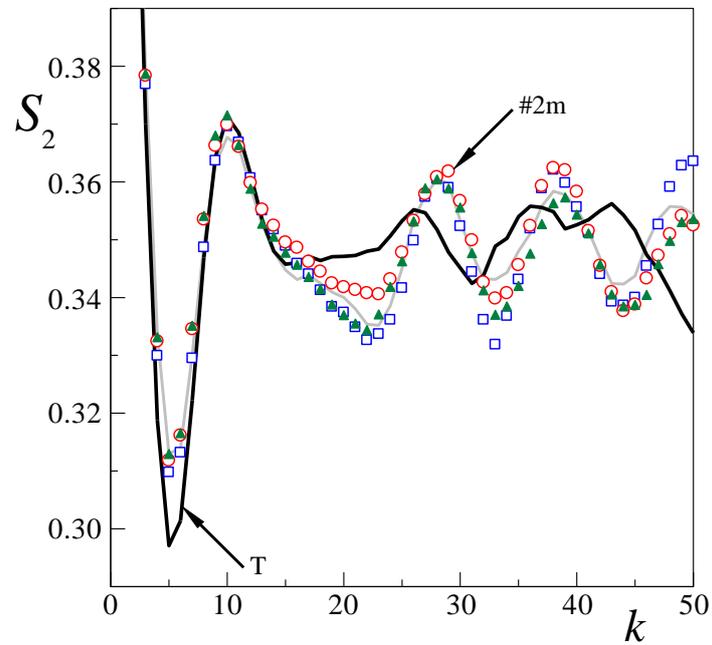

**Figure 3.** Restricted two-point correlation function $S_2$ of black phase computed by the orthogonal-sampling algorithm with the hard-wall conditions for target T-pattern (black line) and for its reconstructions: #2m (red circles), #2b (green triangles), #2rp (blue squares) and #2r (grey line). The easily seen differences between the target and its reconstructions (via EDs) show that they provide relatively dissimilar structural information compared with $S_2$.

Fig. 3